# Synchronized Pivots, A New Way of Engineering Optimized Fc Antibody Fragments

J. C. Phillips

Dept. of Physics and Astronomy, Rutgers University, Piscataway, N. J., 08854

## Abstract

Thermodynamic profiles, based on bioinformatic hydropathicity scales and membrane-fitted sliding windows, exhibit accurately level sets of hydrophobic peaks or valleys in the constant fragment Fc of monoclonal antibodies. These sets provide a roadmap for selected short regions where a few directed amino acid mutations could produce either accelerated antibody kinetics or increased stability. These profiles may reveal dramatic differences in antibody activity. CoV-1 and CoV-2

The immunoglobulin (IgG) superfamily has many remarkable properties, including the ability of IgG conserved fragments Fc to fuse with a wide variety of receptors to produce specific recombinant proteins of enormous medicinal value. Antibody folding and assembly is generally well understood [1], and there are many marketed monoclonal (mAb) products [2], with many more in the pipeline. The amino acid ranges of the conserved heavy chain Fc fragments are: 1-98, CH1; 99-110, disulfide-stabilized hinge at Y vertex; 111-223, CH2; and 224-330, CH3. Numbering is based on Uniprot P01857 (IGHG_1HUMAN).

Many mAb-based drugs utilize all or part of IgG1 or IgG4 CH2 - CH3 humanized fragments containing few or no mutations. These fragments can be recognized not only structurally, but also because they contain a high concentration of evolutionarily conserved amino acid sites [1]. The generic immunoglobulin fold β-barrel structure of antibody domains has been fine-tuned during evolution to optimize both kinetics and stability [1]. Remarkably, most of this fine-tuning carries over to designed monoclonal (mAb) products, but this transferal cannot be optimal. In practice, it has proved simpler to engineer the variable drug component and the variable IgG heavy chain and IgG light chain Fv regions, than the more stable Fc regions. The latter are still important because they stabilize the interactive structure and reduce recombinant product costs.

There is interest in engineering refined Fc fragments, for instance, for extending serum half-life, or even enhancing antibody kinetics. This can be done by increasing Fc binding to protective receptors, possibly by increasing hydrogen bonding through a few mutations in the contact area (although only very small changes were observed in a structural study [3]). This suggests that the observed serum effects involve long-range (conformational) interactions, which are not easily quantified structurally [3,4]. The methods used here are designed to treat long-range interactions quantitatively, and are expected to relate directly to in vivo stability.

The mutations we propose are expected to tune drug kinetics and/or stability, which may well be advantageous, as fusion of humanized Fc to an engineered murinal Fv can slowed antibody activity. At the fundamental molecular level one can see why little effort has been focused on engineering improvements to Fc fragments. The short-range contact interactions of the variable antigen-binding paratopes at the Y tips of IgG are similar to fusing interactions. They are strong and few in number (only billions), and can be studied structurally and statistically with single or pair mutations in Euclidean space. By contrast, the long-range Fc fragment effector interactions are weak and remote, but very large in number, and their effects on kinetics and stability are not easily studied structurally, and even less easily studied with dynamical simulations. Near optimal functionality (criticality), the short-and long-range interactions should be balanced, and equally important, but the long-range (allosteric) interactions of large proteins are not accessible to standard analytical methods.

Information regarding specific drugs (including their FASTA amino acid sequences listed in DrugBank) is often available online, the source used here. The theoretical methods used here have evolved over a period of ten years, and stemmed from a remarkable Brazilian discovery of universal self-similarity in the amino-acid specific hydropathically sculpted surfaces of > 5000 static protein segments in the PDB [5]. The most natural explanation for efficient protein evolution involves modules of length 20-30 amino acids [6]. The Brazilian discovery provided the first solid evidence that modular evolution could be described using fractal parameters based on the statistical mechanics of phase transitions. In such theories thermodynamic scales provide the tools needed for quantitative analysis of modular structures.

This coarse-grained modular approach is radically different from the Newtonian methods used in molecular dynamics simulations. It is well suited to discussing chemical trends in protein superfamilies, such as mAb-based drugs containing fragments of the immunoglobulin (IgG) superfamily. Before embarking on such a large project, one is well-advised to study a few simpler examples of other proteins, in order to gain experience in protein scaling. The author has done so, with encouraging results, as described in several recent reviews [6,7].

The scaling study [6] of mAb-based drugs begins here with the IgG1 heavy chain Fc. The preliminary results focus on the conserved heavy chain fragment Fc, which has been used in many mAb-based drugs. The results of the thermodynamic theory have led to a picture of chain dynamics which has both elastic mechanical and viscous fluid hydrodynamic aspects [6,7]. While the short-range contact interactions usually involve the fused protein and/or the variable chain VH from Fv, the long-range interactions include the Fc supporting fragments (often called effectors).

One can picture the protein action in terms of the hydrodynamic flow around nearly stationary internal maxima (pivots) of the protein profile $\Psi(aa,W)$. Here $\Psi(aa)$ is the specific hydropathicity of the aa site, taken from the 2007 MZ scale [5,6], and W is the wave length of the sliding window used to smooth the short-range interactions to reach the appropriate length scale of flowing modular interactions. When the pivots are nearly level, the flow is better synchronized [6]. Antibodies are strongly interacting membrane proteins, and earlier examples [7] suggested a membrane protein best wave length W = 21. That has proved to be the best wave length for antibodies as well, thereby supporting the view that the present method describes some universal aspects of protein functionality. These are often determined by long-range conformational interactions near a membrane that involve deformations of the water-enveloped protein globule.

In several proteins found in many tissues, protein profile $\Psi(aa,W)$ extrema form level sets for optimized values of W [6]. Fig. 1 shows the results for immunoglobulin IgG1 heavy chain C for human and mouse sequences. The primary human pivots P1 – P5 are almost level (slightly tilted, perhaps by charge interactions), while the primary mouse pivots are merely irregular. The human profile is shown again in Fig. 2, this time labelled for discussion. The irregularity of the mouse profile pivots is consistent with the use of Fc humanized fragments in drugs [1,8,9,10].

The resolution of details achieved by hydropathic profiling is consistent with the latest experiments. In Fig. 3 we compare the profiles of IgG1,2,4 and note the differences. The profiles of IgHG1,4 are similar, with similarly level primary pivots, but differences in the hinge region H1. The hydrophobic pivots of IgHG2 are not level, especially peak P1. This is consistent with maternal antibodies transport. IgG1 is the most efficiently transported subclass and IgG2 the least [11]. Similarly, the hinge regions of IgG1 are more flexible than those of IgG2 and IgG4 [12], in agreement with the deepest IgG1 H1 hinge in Fig. 3.

Level sets appear to be the result of evolution of many proteins towards a nearly optimized configuration with synchronized motions about each pivot [6]. Another test of extremal synchronization involves using the standard hydropathic scale KD [13] for $\Psi(aa)$, which involves protein unfolding (thermodynamically first order), and comparing the results with the modern Brazilian scale (MZ, thermodynamically second order) [5]. The result shown in Fig. 4 is that the level primary pivots found with the modern MZ scale become irregular using the standard KD unfolding scale. Similar results have been obtained for other proteins [6]. They are interpreted as evidence that conformational motion (for instance, binding of antibodies to receptors) involves many combined small distortions of protein globules over large scales. These are not easily recognized structurally [3], and are not well described by first-order unfolding models [14].

Another test of the primary pivot synchronization model for IgG asks how sensitive the level pivot results shown in Fig. 2 with W = 21 are to a 10% change in W. Looking at Fig. 5, we see that the pivots with both W = 19 and W = 23 are qualitatively less level. This implies the existence of a characteristic membrane "frontier" layer for binding antibodies of thickness ~ 21 amino acids [6,7].

The choice W = W* = 21 applies best to membrane proteins. Other proteins can yield different values of W. Hen Egg White is a comparatively small 148 aa protein, which has a nearly centrosymmetric tripartite α helices (1- 56 and 104-148) and β strands (57-103) secondary structure. During its long career, HEW has performed at least three functions, as an enzyme, an antibiotic, and an amyloidosis suppressor. The relative importance of these functions has changed from species to species, and it seems likely that these changes are reflected in the amino acid mutations that have maintained the centrosymmetric structure. Here the value of W which best describes the evolution of HEW is W = 69, roughly half of its length, consistent with its secondary α, β structure [15].

Ubiquitin is a small (only 76 amino acids) universal protein that is responsible for degradation of most proteins: it tags diseased proteins for destruction [16,17]. Remarkably ubiquitin is found unchanged in mammals, birds, fish and even worms. It initiates degradation through a three-stage enzyme cascade. The first E1 stage (UBA_1, Uniprot 22314, 1030 amino acids) retains the same fold and enzyme action from slime mold to human. Profiles of these two proteins with W = 55 (wave length similar to ubiquitin itself) with the thermodynamically second order MZ scale show (Fig. 6) that both have four linearly aligned pivots, with an important difference: the slime mold array is tilted by 15% of the profile width, while the human array is nearly level. Similar profiles using the thermodynamically first order 1982 KD scale show little or no alignment for either profile (Fig. 7). This is another important check on the internal consistency of the scaling method [6]. Such interprotein functional comparisons can be made only with the present modern post-Newtonian hydropathic method.

Many drugs use all or parts of Fc as effectors for support or binding to other proteins. Each drug has larger differences in other components (light chain and variable heavy chain regions, or fused components), and the differences between Fc of IgG (usually IgG1 or IgG4) and Fc in older drugs are usually small (<1%). A faint hint that the hinge region could be engineered occurred in Herceptin (1998), where an extra Proline (which is hydrophilic) is inserted into IgG1at site 101, widening the hinge region. (A BLAST search shows that this strain also occurs in human genes.) The same substitution is used in Keytruda's (2014) BLAST similarities are only 90% with 13 gap sites. Its profile in Fig. 9 shows substantial changes in the hinge region, presumably to achieve greater stability in vivo.

The 21 amino acid epitopic sequence centered on IgG1 primary peak P2 in Fig. 2 is {Z} = 142VTCVVVDVSH*E*DPEVKFNWYV162. This hydrophobic peak occurs in the membrane-bound wave length W = W* = 21. If one chooses a smaller wave length, such as the lower limit of the MZ self-similarity range W = 9, a very different split picture emerges (Fig. 9). The W = 21 peak P2 at 152 is replaced in W = 9 by a dip at 152 between two strongly hydrophobic peaks of $\Psi$(aa,9) = 195 at 145 and 178 at 164, separated by a deep valley of 130 at 122. The P2 coalescent behavior is unusual: peaks 3-5 in Fig. 2 are not split for W = 9, and the separation depth of P1 is four times smaller.

One could imagine many ways to mutate the {Z} epitope to bring P2 ($\Psi$(aa,21) = 170.4) into better alignment ($\Psi$(aa,21) ~ 164) with other pivots. As shown in Fig. 1, the mouse sequence {Z*} = VTCVVVDISKDDPEVQFSWFV reduces P2. Alanining any of the seven Valine sites in {Z} nearly levels P2, and there are many other possibilities. Note that this long-range application of small hydroneutral alanine mutations to {Z} and P2 is qualitatively different from the short-range (W = 1) effects of most alanine mutants [18]. Here strongly hydrophilic glucine (E) is at the center the hydrophobic peak P2 of {Z} = 142VTCVVVDVSH*E*DPEVKFNWYV162, and nearby are aspartic acid (D), another E, and lysine (K). Alaning any of these sites would approximately double the residual height of P2 and increase stability. According to Uniprot, the secondary structure of {Z} is ~ 75% beta strands. Another hydroneutral amino acid is glycine, which is bulky [19] and still might be fitted into {Z}.Other mutations than alanine could be guided statistically by beta strand propensities [6].

There are possibilities for quantifying the misfit strain effects of mutational substitution by using the most advanced molecular dynamics simulation (MDS) tools which show that the water envelope film has an ice-like (not gas-like) structure [20]. Current methods for estimating mutated strain misfit energies are reliable only at high temperatures far above the boiling point of water, and hence far above physiological temperatures [21]. To be effective at lower temperatures, one could combine MDS with the present methods. Thus one could start from a~ 42 amino acid segment centerd on Z and pinned at its ends to an $F_C$ structure [22]. At nanosec intervals one could calculate the solvent accessible surface areas (SASA) of each amino acid site and then compare these $\Psi$(aa,W) and $\Psi$(aa) from [5] for both and WT structures. Comparing these dynamical patterns for a few mutations with experimental data on stability and clonal yield should yield patterns

We can now return to Fig. 1 and ask why the human pivots are much more level overall than the mouse pivots. The obvious explanation is that human antibody fragments Fc have evolved more than mice, and are more effective overall. Another interpretation would be that both BLAST and the self-similar MZ scale are based on human data. If an equally large mouse data base existed, and could be used to define a mouse version of $\Psi$(aa), then with this scale one could find that the mouse pivots were more level than the human. The point of this comment is that the human self-similar MZ scale is extremely accurate; the leveling of the Fc pivots with the MZ scale is not accidental. This point has been made for many other proteins [6], but it is worth repeating here.

An important limitation of crystal structure studies, even of bound complexes, is that they determine only ground states, whereas binding kinetics are determined by transition states. According to [4], "differences between an IgG and its cognate YTE mutant vary with their pH-sensitive dynamics prior to binding FcRn. The conformational dynamics of these two molecules are nearly indistinguishable upon binding FcRn" (here Rn is the much studied neonatal Fc receptor [23,24]. In fact, the YTE mutant region falls close to the P2 peak, but not in the {Z} module. As shown in Fig. 10, the effects of the mutation on the hydropathic profile shown in Fig. 2 are indeed small, and have little effect on the major central P2 peak. A small effect is observed on the secondary peak p*. The effects of the {Z} mutations we have proposed here for reducing the primary peak P2 should be much larger.

In the CH1-CH2-CH3 domains P2 is centrally located, which suggests that evolution has aligned the other pivots in Fig. 2, but increased P2 to provide additional stability, as it did with the central region of Hen Egg White [6]. Stability is desirable for the resting antibody pool, but it also slows kinetics, which may or may not be desirable in a drug. Serum half life of monoclonal antibodies has been the subject of more than 700 articles in the Web of Science. Conventional phage displays scan Fv mutations [25]. The need for long-range Fc synchronization is suggested by cellular studies [26]. Adjusting other peaks also could lead to biobetters [27]. Antibodies engineered to tune through {Z} mutations to prove their stability or long-range kinetics as membrane bound proteins [28] form a promising improvement on present antibodies [29].

Attention is now focusing on the differences between the checkpoint inhibitors Keytruda (Pembrolizumab) and Opdivo (Nivolumab) used in treating advanced melanoma. Their heavy chain profiles are shown in Fig. 11. The differences occur in the N-terminal region of VH and the Hinge region between CH2 and CH3. Keytruda and Opdivo are both Ig4 check point inhibitors. In the VH region, the hydrophobic peak of Opdivo between sites 40 and 60 has become more hydrophobic and shifted to 25-50, closer to the N-terminal. Both differences provide greater stability for Keytruda.  Currently there is little difference in the effectiveness of the two drugs [30].

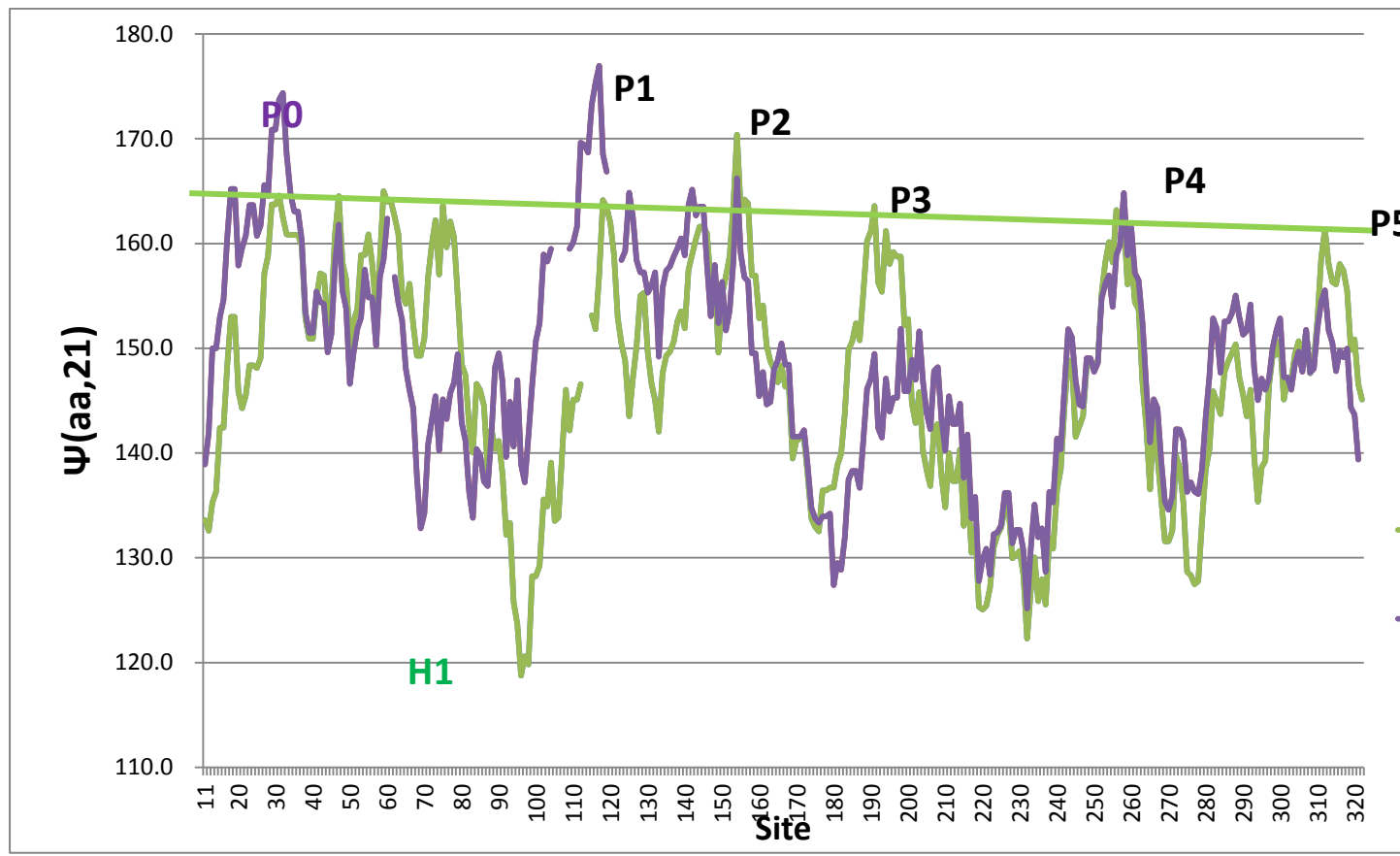

Fig. 1. The hydroprofiles of human (Uniprot P01857) and mouse (Uniprot P01869) IgG1 heavy chain C sequences. The gaps are those given by the BLAST alignment. Here and in all figures, the abscissa is the site number, and the ordinate is $\Psi(aa,W)$ with W = 21. The modern MZ conformational scale is used here, for reasons discussed in text and in [6]. There have been drastic changes in even Fc between human and mouse in two major ways. Human $\Psi(aa,21)$ has developed a single narrow and deep hinge near 94, where mouse had a wider and shallower hinge. The human hydrophobic peaks are nearly level, except for P2, its highest peak. In mouse the peaks are not level, and the highest peak is P1. As we seeof IgGH1 in Fig. 6, the human peak leveling indicates positive Darwinian evolution.

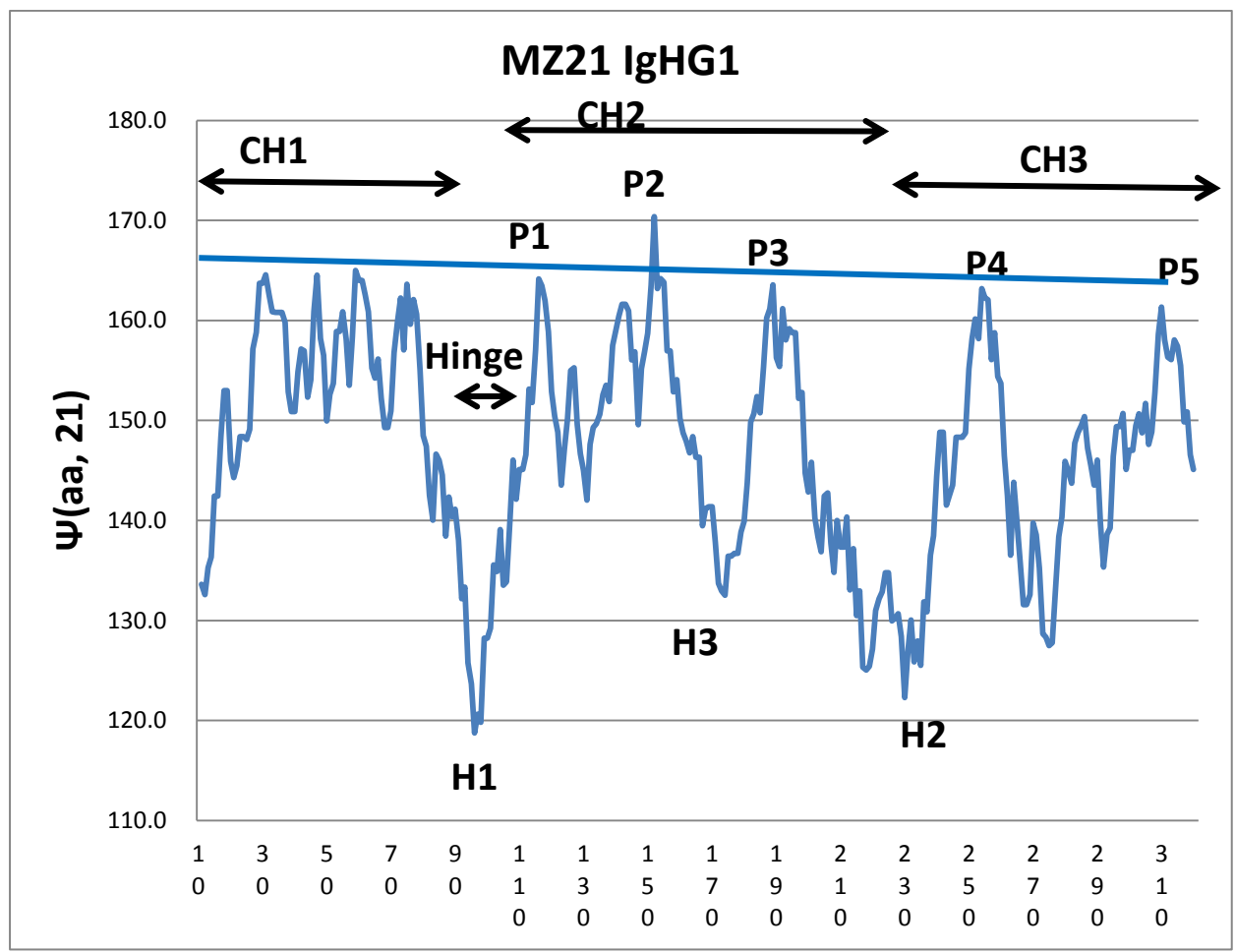

Fig. 2. The human IGHG1 W = 21 profile using the modern MZ conformational scale is annotated using standard structural homologies (Uniprot P01857). [1]. The hydrophobic pivots in CH2 and CH3 are labelled P1 – P5. There are also three hydrophilic hinges H1-H3 of varying depth. In humans H1 is deepest, while in mouse H1 is the least deep (Fig. 1). There are four additional closely spaced pivots in CH1. Nine of the ten pivots are level to 1% with the slightly tilted line, and only P2 is out of line. Peaks P1, P2, …,P5 are centered on 116, 152, 189, 245 and 319.

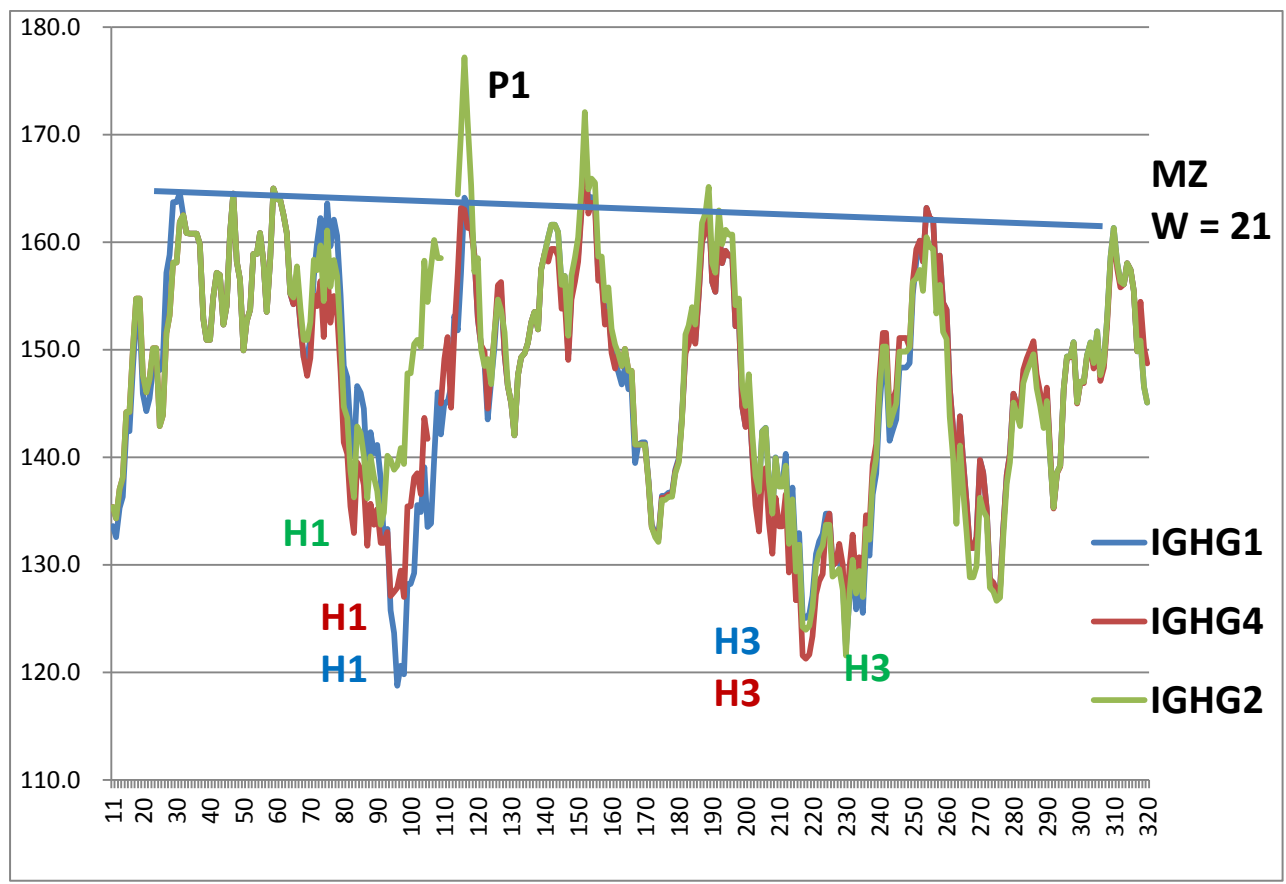

Fig. 3.  The profiles of IgHG1,4 are similar, with similarly level pivots, but differences in H1. The H1 hinge between CH1 and CH2 (see Fig. 4) is 15% deeper in IGHG1 than in IGHG4, which could increase production yield and would explain why IGHG1 is used in older commercial antibodies.  The hydrophobic pivots of IgHG2 are not level, including P2, and especially peak P1, and its H1 hinge is shallow.

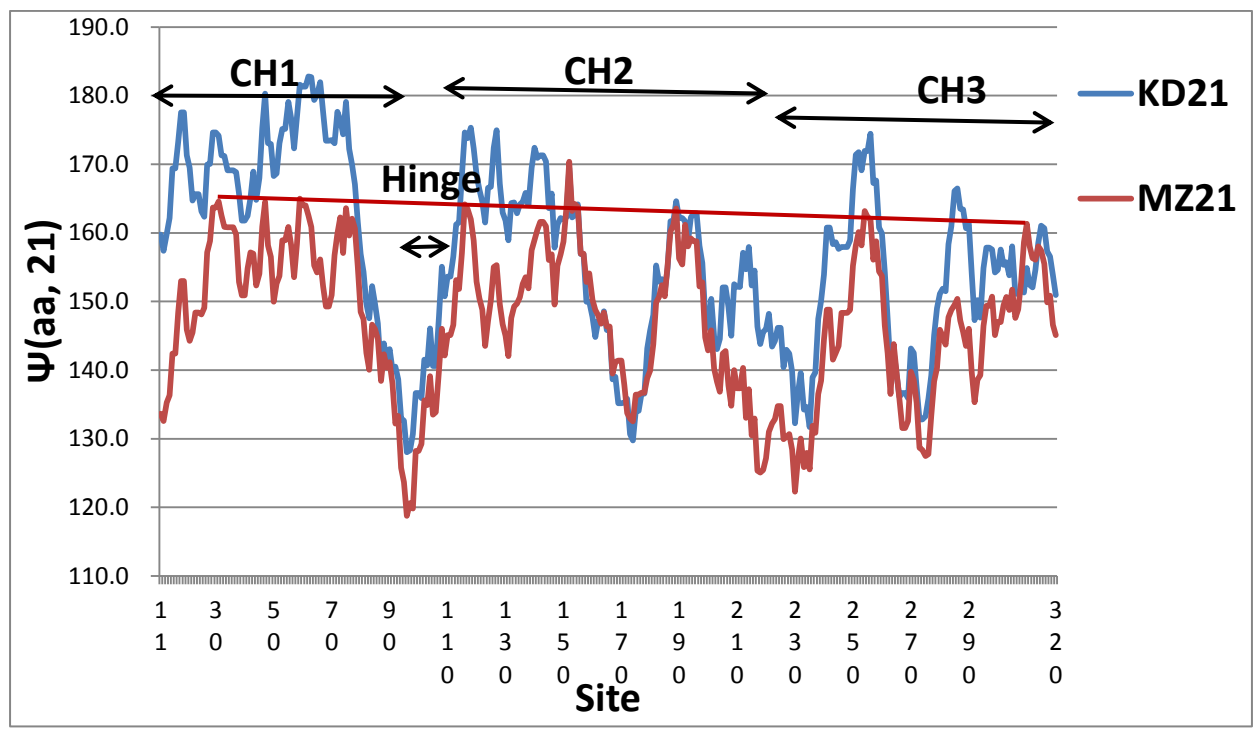

Fig. 4. Comparison of the IgHG1 $\Psi$(aa,W), with W = 21, hydroprofiles using two scales, the conformational MZ scale, and the unfolding KD scale. The MZ conformational pivots are nearly level (except for P2, see Fig. 2), while the pivots are irregular with the KD unfolding scale.

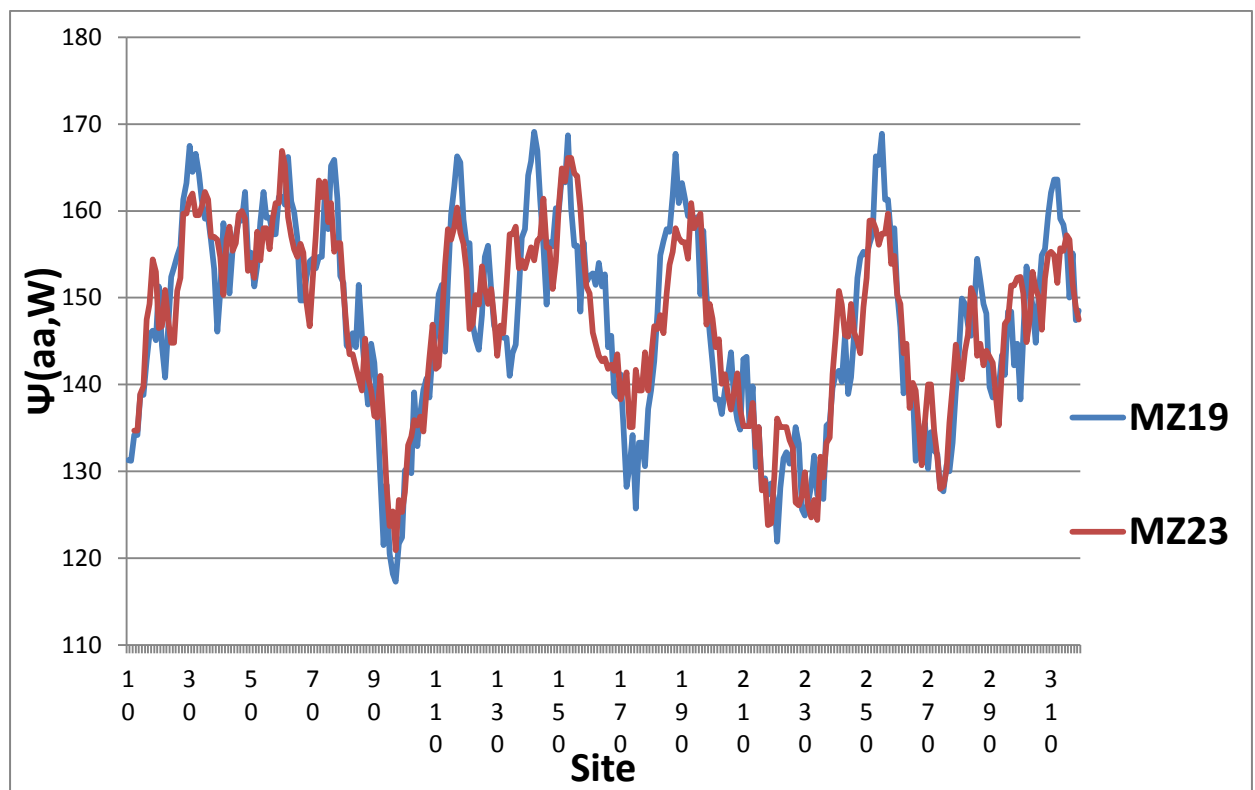

Fig. 5. Comparison of Ψ(aa,W) of IgGH1with W = 19 and 23, using the modern MZ conformational scale. Comparison with Fig. 2 (W = 21) shows that both W =19 and 23, which differ from W = 21 by only 10%, give less level pivots, also by about 10%.

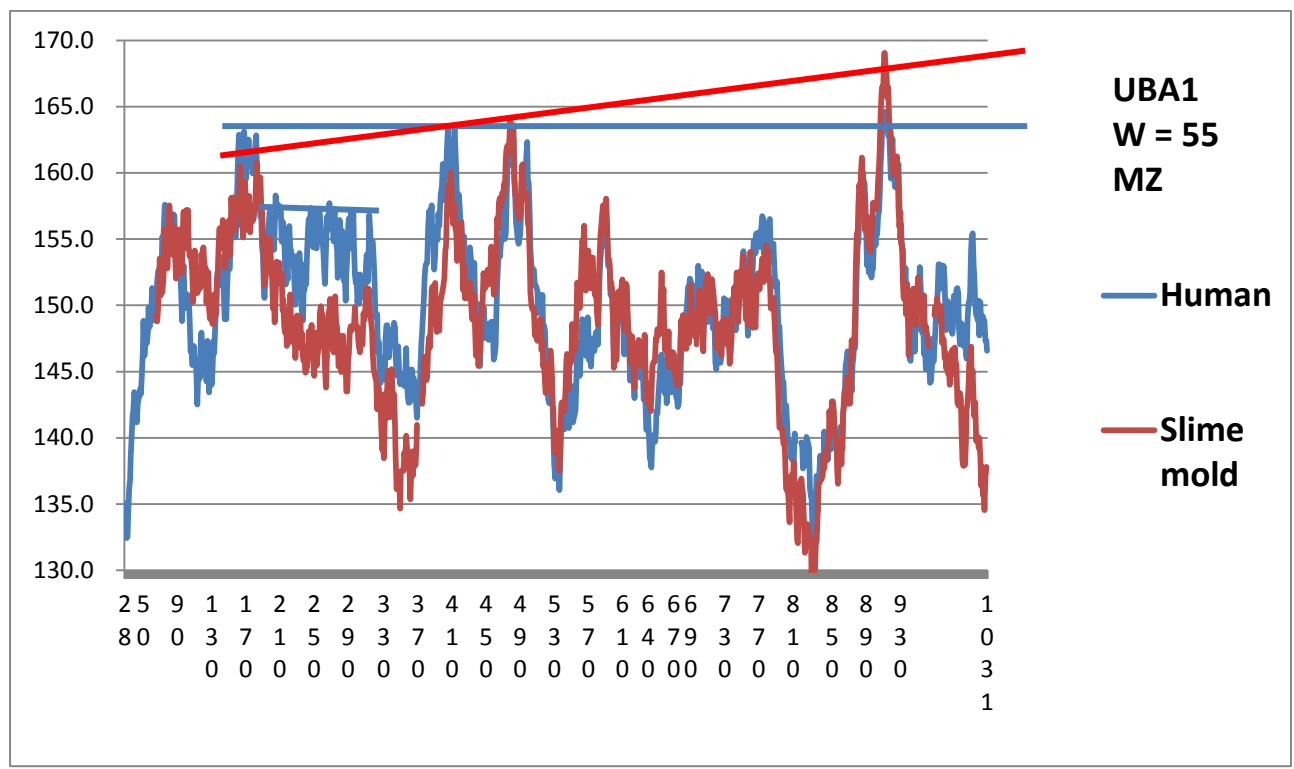

Fig. 6. Comparison of the hydropathic profiles of ubiquitin activating enzyme 1 using the MZ scale. Both profiles show linear pivot alignment, but the human alignment is nearly level. An aligned level subset for human UBA1 is also present between 150 and 290.

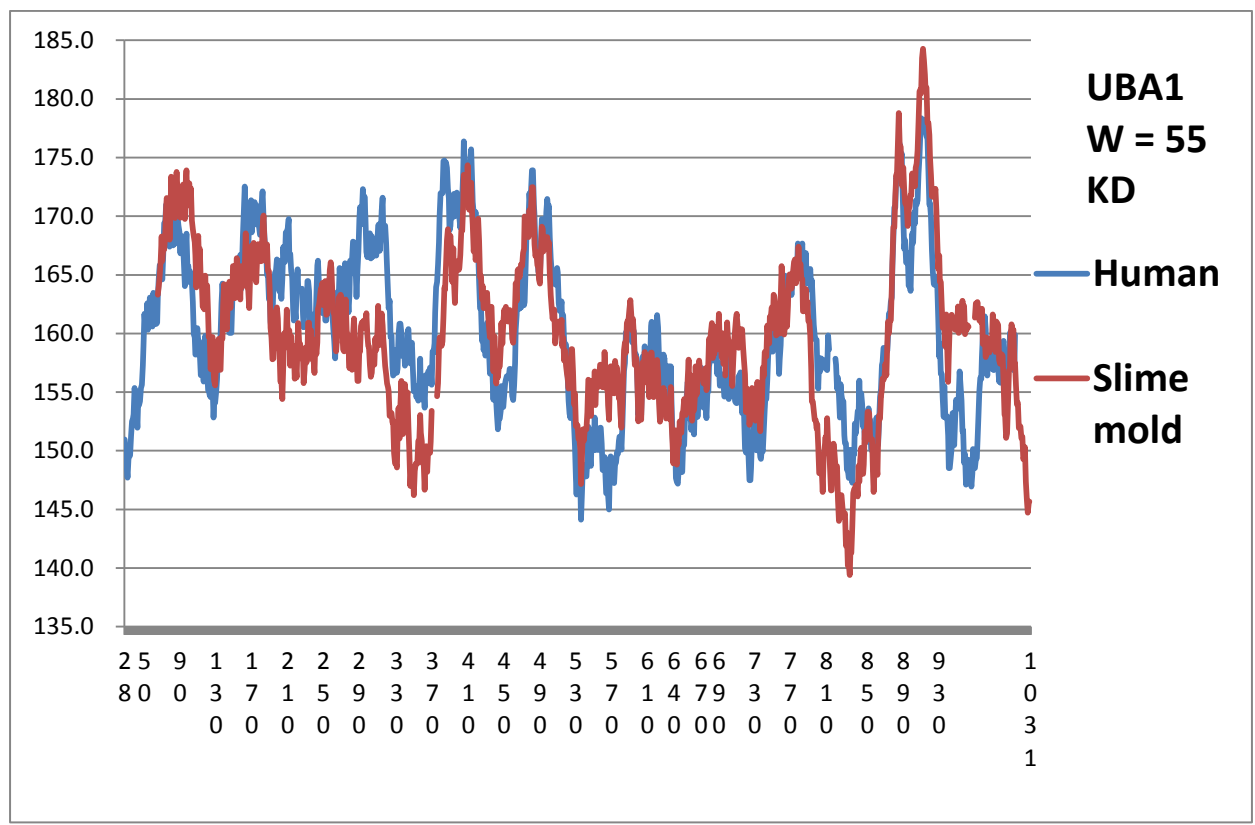

Fig. 7. The UBA1 profiles for human and slime mold using the KD scale. Linear alignment of pivotal maxima is weak here, compared to the excellent alignment seen with the MZ scale in Fig. 6. This means that the interactions of UBA1 with the ubiquitin tag are primarily second order, and have been refined significantly from slime mold to human [6,18]. Compare Figs. 6 and 7 with Fig. 4.

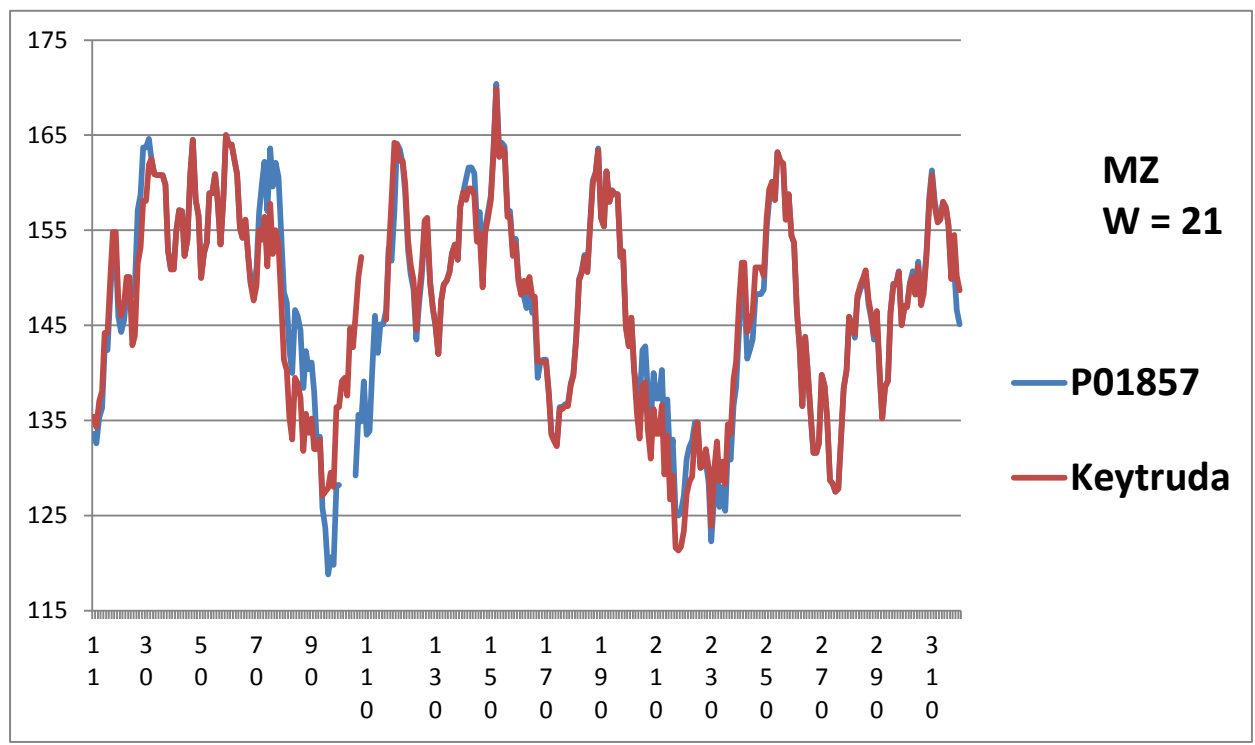

Fig. 9. The largest differences in Ψ(aa,W) with W = 21 between Keytruda and IgG1 are concentrated in the hinge region H1, where Keytruda's depth is less than IgG1's and is similar to IgG4's (Fig. 2).

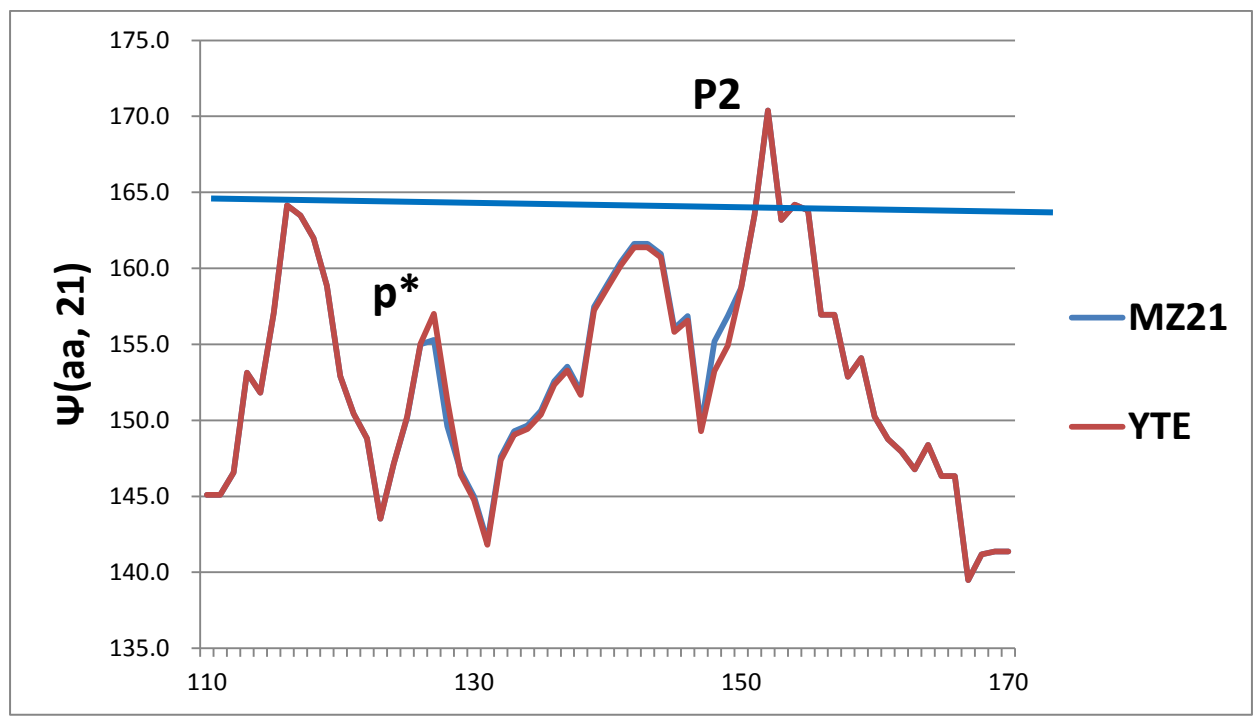

Fig. 10. The human IGHG1 profile using the modern MZ conformational scale is expanded here to show the effects of the much-studied YTE mutation [4]. These are seen to be small and of lesser importance to the conformational dynamics, compared to the excess hydrophobicity of the P2 peak, associated with the {Z} sequence. The YTE mutations assist in stabilizing certain complexes [20].

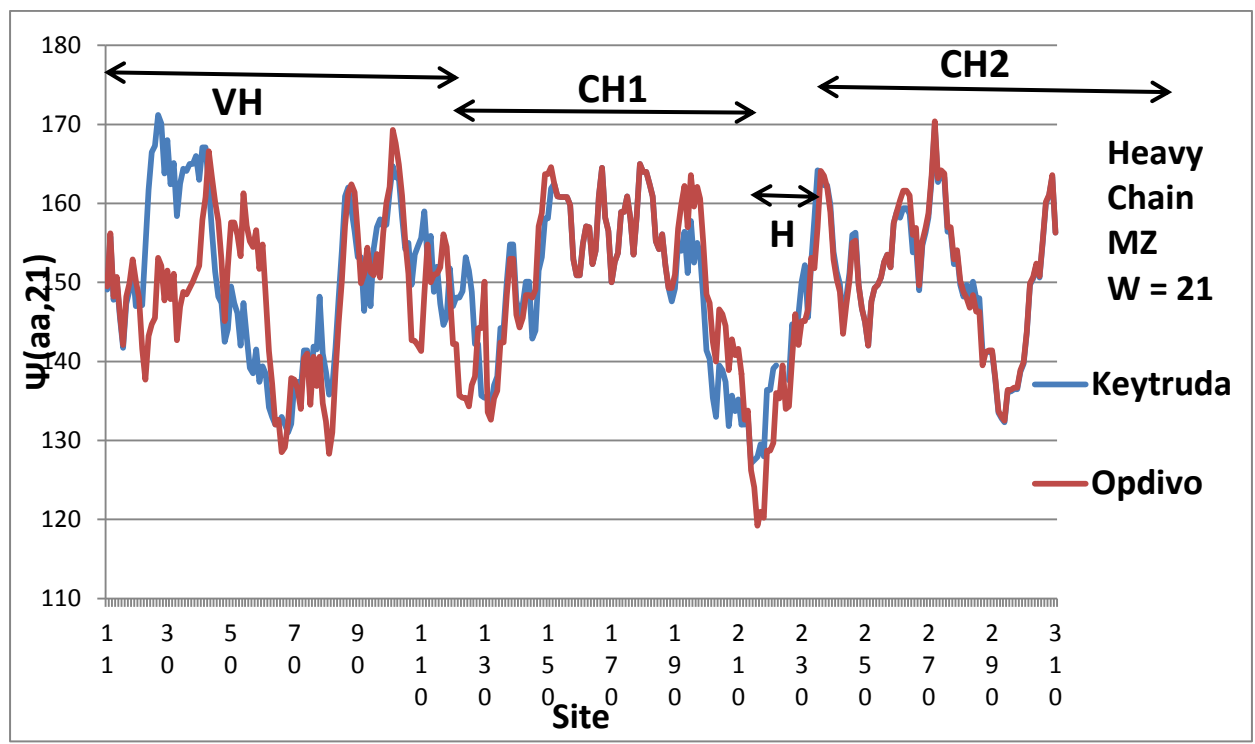

Fig. 11. Comparison of the broadly similar Keytruda and Opdivo heavy chain profiles shows large differences in the N-terminal region of VH. Keytruda is stabilized by a strong hydrophobic peak (sites 20-40)\ and a reinforced hinge H (215).

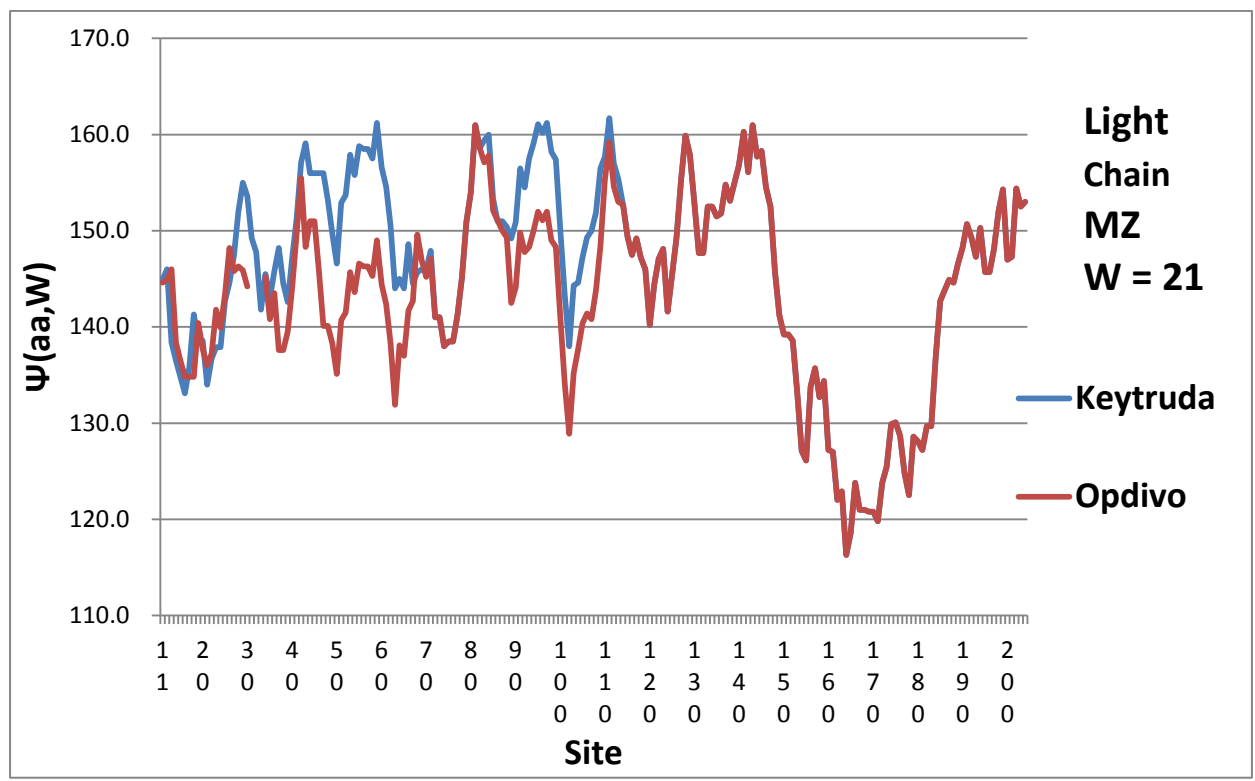

Fig. 12. Comparison of the Keytruda and Opdivo light chain profiles shows large differences in the Complementarity Determining Region, conventionally the 1-40 sites. In the text we suggest that this range, which has been proposed based on static crystal structure contacts, be extended to 100 sites, to include dynamical interactions in transition states. Note also the large level sets of hydrophobic peaks (Opdivo, 81, 111, 119, 143; Keytruda, 59, 81, 97, 111, 143). Keytruda has three level hydrophobic peaks in the 1-100 region, while Opdivo has only one.

`CDR regions 25-33,49-53, 89-96`

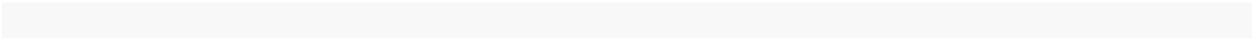